\documentclass[12pt]{article}
\newcommand{\be}{\begin{equation}}
\newcommand{\ee}{\end{equation}}
\newcommand{\els}{e^{-\lambda (t\,-\,s)}}
\newcommand{\elt}{e^{-\lambda t}}
\newcommand{\cts}{\int\limits_s^t}
\newcommand{\sit}{\sigma^2(t)}
\newcommand{\sil}{\frac{\sigma^2}{\lambda}}
\newcommand{\sik}{\sigma^4(t)}
\newcommand{\yks}{u\:-\:v\elt}
\newcommand{\zes}{z^2\:+\:\sik}
\newcommand{\lit}{\lim\limits_{t\to 0}\frac{1}{t}}

\begin{document}
\baselineskip18pt
\oddsidemargin0cm \textwidth15.5cm \textheight21cm

\title{Local characteristics of random motion}

\author{Piotr Garbaczewski\thanks{Email: pgar@proton.if.wsp.zgora.pl}\\
Institute of  Physics, Pedagogical University \\ PL-65 069 Zielona
G\'{o}ra,  Poland \\}

\maketitle

\noindent
\begin{abstract}
Markovian diffusion  processes  yield a system of  conservation
laws which couple various conditional expectation values (local
moments). Solutions  of that closed  system  of deterministic
partial differential equations stand for a regular alternative to
erratic (irregular) sample paths that are associated
with weak solutions of
the primordial stochastic differential equations. We investigate
an issue of local characteristics of motion in the non-Gaussian
context, when moments of the probability  measure may not exist. A
particular emphasis is put on jump-type stochastic processes with
the Ornstein-Uhlenbeck-Cauchy process as a fully computable
exemplary case.

\end{abstract}

\section{Local characteristics of the Brownian motion}

Lets us consider a Markov process $X_t$ in $R^1$. We can fully
characterize an associated random dynamics by means of  a
transition density $p(y,s,x,t)$ with $0\leq s < t$
and an initial density $\rho _0(x)=\rho (x,t_0), 0\leq t_0$.
For the stochastic process to be properly defined we
impose the so-called stochastic continuity condition:

\be lim_{t\downarrow s} {1\over {t-s}} \int_{|y-x|> \epsilon }
p(y,s,x,t) dx = 0 \ee
to be valid for arbitrary (every) $\epsilon >0$ and  for almost
every $y\in R^1$. That is known to imply that along a sample path
$\omega $ there holds  $lim_{t\downarrow s} P[\omega : |X_t(\omega
) - X_s(\omega )| \geq \epsilon ] = 0 $.
This condition needs  be respected by both diffusion and jump-type
propagation scenarios and is known to be necessary but still insufficient
for the Markov process to have continuous trajectories.

Various characteristics of random motion can be directly associated with
properties of sample trajectories (consider for concretness the
iterated logarithm laws in the Brownian motion).
Those characteristics of random  dynamics, that are free of
irregularities of sample paths, need to involve some forms of
 averaging. That refers either  to global averaging present
while evaluating moments of a probability measure (provided they
exist)  or to the evaluation of local moments of  that  measure.

Given a transition probability of a Markov process, we can attempt to derive
a number of local expectation values, like e.g. the forward drift of the process:

\be b(x,s)=lim_{t\downarrow s} {1\over {t-s}} \int_{|y-x|\leq
\delta } (y-x) p(x,s,y,t) dy \ee
and the diffusion (coefficient if a constant)  function

\be D(x,s) = lim_{t\downarrow s} {1\over {t-s}} \int_{|y-x|\leq
\delta } (y-x)^2 p(x,s,y,t) dy \ee
where the $\delta $ cutoff is  needed to guarantee   a convergence of the
integral. In principle, the drift
and diffusion functions should take values independent of  the
$\delta $-cutoff and/or exist when the cutoff is removed (once  we let
$\delta $ go to $\infty $).
 However, for the jump-type processes, local moments (e.g.
drifts and diffusion functions) in the absence of a cutoff are
generically nonexistent, while in a cutoff version they
display an explicit and nontrivial $\delta $-dependence.

As yet, our discussion extends to both continuous and discontinuous
processes.  Let us explicitly reveal when a diffusion process
enters the game.
By assuming that  there holds:
\be lim_{t\downarrow s} {1\over {t-s}} \int_R |y-x|^{2+\gamma }
p(x,s,y,t) dy =0 \ee
for any $\gamma >0$, we set a sufficient condition  for the
continuity of sample paths (to hold true almost surely), that
also allows  to remove the previous $\delta $-cutoff from local
momet formulas. The resulting process is  a Markovian
diffusion process with the forward drift $b(x,s)$ and diffusion
function $D(x,s)$.
Our futher discussion will  be carried  under a
simplifying  assumption that a diffusion
function  $D(x,s)$ actually is a constant which we denote $D$.

Our previous considerations, when specialized to Markovian
diffusion-type processes, can be cast in  another form. Namely, we
can depart from a formal infinitesimal version of a stochastic
differential equation for a random variable $X_t=X(t)$ taking values
in $R^1$:

\be dX_t= b(X(t),t)dt + \sqrt{2D} dW_t
\ee

where $W_t=W(t)$ is  a Wiener process  and $b(x,t)$ stands for the forward drift of
the diffusion-type process $X_t$.

If we assign
 a probability density $\rho _0(x)$ with which the
 initial data  $x_0=X(0)$ for the stochastic differential equation  are
distributed (weak solutions  enter the scene), then
the emergent Fick law  would  reveal a
statistical tendency of particles to flow away from
higher probability  residence areas.     This
feature is encoded in the corresponding Fokker-Planck equation
(equivalently, a continuity  equation):
\be \partial _t \rho  = - \nabla \cdot (v\rho )\ee
where a diffusion current velocity  is
\be v(x,t) = b(x,t) - D{{\nabla \rho
(x,t)}\over {\rho (x,t)}}\, .\ee

Clearly, the local diffusion current (a local flow that might
be experimentally observed for  a cloud of suspended particles
in a liquid)
$j=v \rho $ is
nonzero  in the nonequilibrium  situation and quantifies
a non-negligible  matter transport which occurs as a
consequence of  the Brownian motion, on the ensemble average.

It is interesting to notice that the local velocity field
 $v(x,t)$
  obeys the natural (local)  conservation law, which we
quite intentionally pattern after the  moment identities
(hierarchy of conservation laws) valid for the Boltzmann and Kramers
equations.
 The pertinent momentum conservation law  directly originates from
the rules of the It\^{o} calculus for Markovian diffusion processes,
 and from the first moment equation in the
diffusion approximation  of the Kramers theory:
\be \partial _tv + (v \cdot \nabla ) v =
\nabla (\Omega - Q)\enspace .\ee

 The  general
 form of an auxiliary potential $\Omega (x,t)$
reads: \be \Omega (x,t) = 2D[ \partial _t\phi + {1\over 2}
({b^2\over {2D}} + \nabla \cdot b)]\enspace .\ee and  can be
interpreted as a constraint equation for the admissible forward
drift $b(x,t) = 2D \nabla  \phi (x,t)$. It is useful to mention
that the above  momentum conservation law may be regarded to
account for effects of external (conservative) volume forces
$-\nabla (-\Omega )$ which are determined  by a priori arbitrary,
but  bounded from below, continuous function $\Omega (x,t)$.
Then,  the allowed forward drift must be disentangled from the
Ricatti-type equation (9) as a function of  $\Omega $.

In the above
 there appears a contribution from a
  probability density  $\rho $-dependent potential
 $Q(x,t)$. It is  is given in terms of the so-called osmotic
velocity field  $u(x,t)$:
\be Q(x,t) = {1\over 2} u^2 + D\nabla
\cdot u\ee
$$u(x,t) = D\nabla  \, ln \rho (x,t)$$
and  is  generic to a local momentum conservation
law  respected by   isothermal Markovian diffusion processes.
 Notice that in  case of the free Brownian motion (admitted, if
 we set $\Omega = 0$), we would have
 $v(x,t) = - u(x,t)$ for all times.
 An equivalent form of the pressure-type potential $Q$ is
 $Q= 2D^2{{\triangle \rho ^{1/2}} \over {\rho ^{1/2}}}$.

It is interesting to observe that a gradient field ansatz for the
diffusion current velocity ($v=\nabla S$): \be \partial _t\rho =
- \nabla \cdot [(\nabla S)\rho ] \ee
 allows to transform the momentum
conservation law  of a Markovian diffusion process to  the
universal Hamilton-Jacobi form:
\be \Omega =
\partial _tS + {1\over 2} |\nabla S|^2  + Q  \ee
where $Q(x,t)$ was defined before. By applying
the  gradient operation we immediately recover the previous
local momentum conservation law.

Notice that the above Hamilton-Jacobi- type equation  is sensitive to
any additive (constant or
time-dependent) modification of the potential $\Omega $. In the
above,  the contribution due to $Q$ is a direct consequence of  an
initial probability measure choice for the diffusion    process,
 while $\Omega $  alone does account for an appropriate
forward drift of the process.

The simplest realisation of  the outlined  theoretical framework
can be provided by invoking a standard Ornstein-Uhlenbeck process.
Namely, let us consider an It\^{o} equation (in  its symbolic differential
version) for infinitesimal increments of the velocity random
variable, exhibiting the systematic frictional resistance:
\be dV(t)=-\beta V(t)dt + \beta \sqrt{2D} dW(t) \ee
where $W(t)$ is  the normalized Wiener process.
One can  easily infer  the corresponding second
Kolmogorov (Fokker-Planck) equation
\be \partial _t p(v_0,v,t) = \beta D \triangle _{v}
+ \beta \nabla _{v}\cdot [v\,  p(v_0,v,t)]
\ee
for the transition probability density of the time homogeneous
process in the velocity space alone.
The pertinent transition probability density reads:
\be p(v_0,v,t) =
({m\over {2\pi kT(1- e^{-2\beta t)}}})^{1/2}
exp \{ {m\over {2kT}} {{(v - v_0e^{-\beta t})^2}
\over {1-e^{-\beta t}}}\} \ee

Let us consider an instantaneous velocity  $V_t=v$,
 that  has been achieved in the course of the  Ornstein-Uhlenbeck
 random evolution
 beginning from a  certain $V_0=v_0$.
We can evaluate a conditional expectation value (local mean with
 respect to the law of random displacements) over all randomly
 accessible velocities   $V(t+\triangle t)= v'$
at a  time  $t+\triangle t,\,  \triangle t>0 $. That  determines
the  forward drift of the process:
\be b(v,t)= lim_{\triangle t \downarrow 0}
[\int v'\,
p(v,v',\triangle t)\, dv' - v] = -\beta v
\ee
and  thus   provides us with an information
about the  mean tendency
of the dynamics  on  small  (but not too small if compared to
 the relaxation time $\beta ^{-1}$)  scales.
Analogously, we can derive a diffusion function for the
Ornstein-Uhlenbeck process which  is constant and equals
 $2D\beta ^2$.

To arrive at local conservation laws, an additional input of an
initial probability density $\rho _0(v)$ of $V_0$ is necessary. To
that end, one may choose an asymptotic stationary (invariant,
Maxwell-Boltzmann) density of the Ornstein-Uhlenbeck process:
$\rho _0(v) = ({m\over {2\pi kT}})^{1/2} exp[{{mv^2}\over
{2kT}}]$.

\section{From Poisson probability law to  jump - type processes}

A random variable  $X$ taking discrete values $0,y,2y,3y,...$,
with $y>0$ is said to have Poisson distribution ${\cal P}(\lambda
),\, \lambda \geq 0$ with jump size $y$, if  the probability of
$X=ky$ is given by $P(X=ky)= {{\lambda ^k}\over {k!}}exp(-\lambda
)$. The characteristic function of ${\cal P}(\lambda )$ reads:
\be
{E[exp(ipX)] = exp[\lambda (e^{ipy} - 1)] = \Sigma _0^{\infty }
e^{-\lambda }\, {{\lambda ^k}\over {k!}}e^{ik(py)}\, =\, \Sigma
_0^{\infty } e^{ik(py)}\, P(X=ky)}\ee
 and  the  first moment of the probability measure equals
$E[X] = \lambda $. Notice that $P(X=0)= exp(-\lambda )$, hence the
numerical value of $\lambda \geq 0$ fixes the no-jump probability.
 For the Poisson random variable with values $b+ky, k=0,1,...$we
 would get
${E[exp(ipX)]= exp[ibp +\lambda (e^{ipy} - 1)]}$.

 If we consider  $n$ independent random  variables $X_j,
1\leq j\leq n$ such that  $X_j$ has Poisson distribution ${\cal
P}(\lambda _j)$ with jump size $y_j$, then a new variable $X$ can
be introduced by means of the  distribution of $X_1+...+X_n$ whose
 characteristic function reads
 \be {E[exp(ipX)]= exp[\Sigma
_{j=1}^n \lambda _j(e^{ipy_j} - 1)]}. \ee

The exponent in the above might  include an additional term
$ip\Sigma _1^n b_j$ if nonrandom shifts of each jump $ky_j$ by
$b_j$ were allowed.

We can admit not only jumps of fixed magnitudes $y_1,...,y_n$ but
also jumps covering an arbitrary range in $R_+$. Let the
distribution function of the magnitude of the jump be $P(x<y)=\mu
(y)$.  In this case we set \be  {E[exp(ipX)]= exp[\int_{R_+}
(e^{ipy} - 1) \mu (dy)]}\ee
 assuming that the integral in the
exponent exists. (Notice that the previous formula  is recovered,
if we choose ${\mu (dy)=\Sigma_{j=1}^n \lambda _j\delta (y-y_j)\,
dy}$.)

For  any Borel set $A\subset R $ bounded
away from the origin, the random variable  $X_A$ representing  jumps bounded by $A$,
gives rise to a  characteristic exponent $\int_A (e^{ipy}-1)d\mu (y)$, and
the expected number $E_A[X]$ of jumps of size bounded by $A$ is
equal to $\mu (A)$. We can interpret that in terms of jumps of
different sizes that are mutually  independent. Jumps whose size is
bounded by $[y,y+\triangle y), \, \triangle y\ll 1$, do contribute a
Poisson component with exponent function approximately equal to
$(e^{ipy}-1)\mu ([y,y+\triangle y))$.

Let  us consider  an expression for a characteristic function of a
probability measure of a certain random variable $X$ that is given
in the form $E[exp(ipX)] = exp[- F(p)]$ where $p\in R^1$  and for
$-\infty <p<+\infty $, $F=F(p)$ is a real valued, bounded from
below, locally integrable function. If $F(p)$ satisfies the
 celebrated L\'{e}vy-Khintchine formula, then the pertinent  measure
 is positive  and we may introduce positivity preserving
 semigroups, together with the induced (Markovian) stochastic processes.
Let us concentrate our attention on  the integral part of the
L\'{e}vy-Khintchine formula, which is responsible for arbitrary
stochastic jump features. In that case,   $F(p)$ takes the form:
\be
{F(p) = - \int_{-\infty }^{+\infty } [exp(ipy) - 1 - {ipy\over
{1+y^2}}] \nu (dy)}\ee
 where $\nu (dy)$ stands for
the so-called L\'{e}vy measure on $R^1$.

A generic feature of jump-type processes is that they admit jumps
of arbitrarily small size (without any lower bound) and some care is necessary when
evaluating contributions from a close neighbourhood of the
origin.   One obvious way to bypass this  problem amounts to an
$\epsilon $-cutoff which allows to neglect "too small" jumps. Namely,
 for a probability law with the characteristic exponent
$-F(p)$, we can consider its restriction to upward jumps of size
exceeding a given lower bound  like e.g.  to all $y>\epsilon
>0$:
\be {\phi _{\epsilon }^+(p) = \int_{y>\epsilon } [e^{ipy}-1-
{{ipy} \over {1+y^2}}]\nu (dy) = \int_{y>\epsilon }[e^{ipy} - 1
]d\nu (y)\, -\, ipb _{\epsilon }^+}\ee
 $$b_{\epsilon }^+ =
\int_{y>\epsilon } {y\over {1+y^2}}\nu (dy).$$

 Clearly,  we deal
here with a random variable  of the type considered before, and
we can  try to isolate contributions from jumps  of the size
$[y,y+\triangle y)$ by  coarse-graining a Borel set $A$ of
interest. A formal exploitation of
\be {\nu (dy)=\Sigma_{j=1}^n
\lambda _j\delta (y-y_j)\, dy}\ee
 gives rise to
 \be
{E[exp(ipX)]= exp[\Sigma _{j=1}^n [\lambda _j(e^{ipy_j}-1) - ip\,
{{\lambda _jy_j}\over {1+y^2_j}}]}.\ee

Further specializing the problem  we  shall consider
 L\'{e}vy measures that obey the spatial reflection property
$\nu (-dy)=-\nu (dy)$. Then,  we can readily
extend our discussion to jumps of all sorts  in $R^1$, i.e. $y$ can
take values in both $R_+$ and $R_-$, with the only restriction to
be observed that $|y|>\epsilon >0$.
 Notice that we shall deal with
two distinct types of jumps,   either positive or negative, with
no common jump point for them. This
fact means that they are independent components of the more
general random variable:
\be {\phi _{\epsilon }(p) =
\int_{|y|>\epsilon } [e^{ipy} - 1]\nu (dy) - ipb_{\epsilon
}}\ee
 where (choose $\nu (dy)={{dy}\over {\pi y^2}}$ for concretness)
 the deterministic term identically vanishes in view of
\be {b_{\epsilon }= b_{\epsilon }^+\, +\, b_{\epsilon }^- =
\int_{y>\epsilon }{y\over {1+y^2}}\nu (dy) \, +\,
\int_{y<-\epsilon } {y\over {1+y^2}}\nu (dy) \equiv 0}. \ee

In the  previous steps we have indirectly exploited a defining  property of
infinitely divisible probability laws:   if $exp\phi (p)$ is a characteristic
function of a given probability distribution, then
$[exp\phi (p)]^t = exp[t\phi (p)],\, t>0$ is likewise a characteristic
function of an infinitely divisible probability law again. This
feature  extends our discussion to
stochastic jump and jump-type processes (time homogeneous with independent
increments).
Obviously,
for such processes $E[exp(ipX(t))]=exp[t\phi (p)]$ while
$E_A[X(t)]=t\nu (A)$,  and our previous arguments retain their
validity with respect to
\be {E[exp(ipX(t))]_{\epsilon } = exp[t\phi
_{\epsilon }(p)] = exp[t\, \int_{|y|>\epsilon }(e^{ipy}-1)\nu
(dy)]}. \ee

Coming back to the  L\'{e}vy-Khintchine formula, let us replace a
function $F(p)$ by an operator acting in a suitable domain
according to the recipe: $F(p) \rightarrow
\hat{H}= F(\hat{p})$  where $\hat{p} = - i\nabla $.
We easily learn that for times $t\geq 0$ there holds
\be {[exp(-t\hat{H)}]f(x) =  [exp(-tF(p)) \hat{f}(p)]^{\vee }(x)}\ee
where the superscript $\vee $ denotes the inverse Fourier
transform and
$\hat{f}$  stands for the Fourier transform of a function $f$.

If we  set $p_t={1\over {\sqrt {2\pi }}}[exp(-tF(p)]^{\vee }$, then
 the action of $exp(-t\hat{H})$ can be given in terms of a
convolution: $exp(-t\hat{H})f = f*p_t$, where $(f*g)(x): =\int_R
g(x-z)f(z)dz $.
Clearly, there holds:
\be \partial _t\rho (x,t) = - (\hat{H} \rho )(x,t)\Longrightarrow
\ee

$$\partial \rho _{\epsilon } (A,t) = \int _Rdx[\int_{|y|>\epsilon }
[\chi _A(x+y) - \chi _A(x)]\nu (dy)] \rho _{\epsilon }(x,t) $$
which displays a generic Master-equation form.
Indeed, we have here $\partial _t\rho _{\epsilon }(A,t) = \int_R
q_{\epsilon } (x,t,A) \rho _{\epsilon }(x,t) dx$ where
$q_{\epsilon }$ is interpreted as the jump intensity.

Let us however emphasize that the above simplification
occurs only in the $|y|>\epsilon>0$ jumping size regime. The real
r\^{o}le of two integral terms in the expression for $b_{\epsilon }$
is to compensate the divergent
contributions from the L\'{e}vy measure when the principal value
integral $\epsilon \rightarrow 0$ limit is considered; then the
\it standard \rm jump  process theory  does not literally apply
since arbitrarily small jumps are admitted.
Anyway, those two terms  are irrelevant if we assume an  $\epsilon >0$ cutoff,
irrespective  of  how small  the chosen (and fixed)  $\epsilon $ is.

The best known example of the stable probability law  that is
compatible with the above definitions is provided by the classic
Cauchy density which will be our reference  probability law in below.
Let us
focus our attention on that case which is   specified by    $ F(p)= |p|$.
The corresponding semigroup
generator  $\hat{H}= |\nabla |$
is a pseudodifferential operator.
The associated kernel $p_t$  is a transition density of
the jump-type Cauchy  process, which is a solution of
 a pseudodifferential  Fokker-Planck equation:
\be \partial _t {\rho
}(x,t)=
 - |\nabla|{{\rho }}(x,t).\ee

The pertinent   probability density reads  $\rho
 (x,t)={1\over \pi }{t\over {t^2+x^2}}$ and the corresponding
 space-time homogeneous transition  density (e.g.
 the semigroup kernel function) is:
 \be \rho (x,t) = {1\over {\pi }}\, {t\over {t^2 +
x^2}} \Longrightarrow p(y,s,x,t) = {1\over {\pi }}{{t-s}\over
{(t-s)^2 + (x-y)^2}}]\ee
 $$0<s<t$$
 $$\langle exp[ipX(t)]\rangle
:= \int_R exp(ipx) \rho (x,t)\, dx = exp(-|p|t).$$

The characteristic function of $p(y,s,x,t)$ for $y,s$ fixed, reads
$exp[ipy - |p|(t-s)]$, and the L\'{e}vy measure needed to evaluate
the L\'{e}vy-Khintchine integral reads: ${\nu (dy): =
lim_{t\downarrow 0} [{1\over t}p(0,0,y,t)]dy ={{dy} \over {\pi
y^2}}}$.

The Cauchy process   belongs to the category of
jump-type processes,
where apart from  the long jumps-tail (no fixed bound can be
imposed on their length) which is the reason of the nonexistence of
moments of the probability measure, sample paths of the
 process may have an infinite number of jumps of arbitrarily small size.
By general arguments,
pertaining to the space $D_E[0,\infty )$ of right continuous
functions with left limits (cadlag),  both in the finite and
infinite time interval the number of jumps is at most countable.

An introduction of the $\epsilon $-cutoff which eliminates  small jumps
is fed up by the physical intuition. An  approximation of the
jump-type process should in principle be possible   in terms of more traditional
jump  processes which involve a finite number of jumps in a finite time
 interval.
This radical approximation assumption is usually achieved by
 giving a characterisation of the affiliated Markovian jump-type processes
in terms of approximating  (convergent)  families of so-called
\it step \rm processes. The step processes are not yet  the
  jump processes of the standard daily experimental evidence.
They need to have  no accumulation points of jumps in a finite
time interval: in that case the number of jumps is finite on
each finite time interval and between jumps the sample path is constant.

Let us recall that the operator $|\nabla |$ acts as follows:
\be |\nabla |f(x) = - {1\over \pi }\int_R [f(x+y) - f(x) -
{{y\nabla f(x)}\over {1+y^2}}] {dy\over y^2}\, .\ee

By turning back to the pseudodifferential Fokker-Planck equation
with the $\epsilon $-cutoff implemented, let
us introduce  an  operator $|\nabla |_\epsilon $:
\be {|\nabla |_\epsilon f(x)= - {1\over \pi } \int_{|y|>\epsilon }
[f(x+y) - f(x)]{dy\over y^2}}\, . \ee
It suffices to replace $f$ by $\rho _{\epsilon }$ to arrive at the
right-hand-side of the previously defined Fokker-Planck equation
for the step process
approximant of the Cauchy process. Its generator is
just $|\nabla |_\epsilon $.

While mitigating the "arbitrarily small jumps" problem the
$\epsilon $-cutoff does not remove all obstacles related to the
Cauchy process. Indeed, at the first glance the situation looks
deceivingly simple, because on  a finite time interval there can
be at most finitely many points $t\in [0,T]$ at which the jump
size exceeds a given positive number. In view of that,  $sup_{t\in
[0,T]}\, |X_t^\epsilon | < \infty $ where  $X_t^\epsilon $ stands
for the $\epsilon $-bounded Cauchy process (the same argument
extends to the unrestricted Cauchy process $X_t$). However, there
must be \it no   fixed \rm upper bound for the size of jumps
(except for being finite), since  a stochastically continuous
process with independent increments having, with probability 1, no
jumps of size exceeding a certain constant $\delta $, would
possess all moments. That is certainly not the case for the
Cauchy process, which is known   not to have any moments.

Hence imposing or not imposing an upper bound on the jump size (call it a
$\delta  $ cutoff) is another critical issue that hampers a
reliable approximation of the jump - type process in terms of
experimentally verifiable jump processes whose jump size is  bounded
both from below and from above. That derives  in  part from to the resolution
limitations  of realistic experimental arrangements (any experimental data
collection and any  computer simulation/experimentation have built-in
lower and upper  jump size bounds), and in part from
 the fact that all observations are carried in finite time on
 systems of finite spatial extension.
Let us point out that by imposing both  $\epsilon $ and $\delta $
cutoffs on the Cauchy process, we would reduce the problem to the
standard jump process.

Obstacles related to heavy tails of the probability distribution
can be visualized by making computer experiments for the
converging (in fact, diverging) variance test. Namely, having
given a sample of jumps determined by the Cauchy distribution:
$X_1, X_2,...X_n$, with $1\leq i\leq n$,one can form a statistics
based on first $i$ "observations" and ask for the  behaviour of
the averages with respect to $i$. First we need the $i$-th mean
value: $ \overline{X}= {1\over i}\sum_{k=1}^i X_i $
 and next the $i$-th variance:
 $ S^2_i=
{1\over {i-1}} \sum_{k=1}^i (X_k - \overline{X}_i)^2 $.
Plots of $\overline{X}_i$ and $S^2_i$ against $i$ would show up a
fairly irregular behaviour and definite \it non-convergence  \rm
signatures for large values of $i\leq n$.

\section{Local characteristics of the Ornstein-Uhlenbeck-Cauchy
process}

Although we directly refer to the specific Cauchy stochastic process, in
fact we stay in a much broader  setting of so-called  L\'{e}vy flights.
Generically, the variance and higher cumulants of those processes
are infinite (nonexistent). There is also  physically more
singular subclass of such processes for which even the
 first moment (mean value) is nonexistent. That is true for the
 Cauchy process.
Thus we need to relax the limitations of  the standard Gaussian
paradigm:  we  face here  a fundamental problem of
establishing   other means (than variances and mean values)
to characterize statistical properties
of L\'{e}vy processes.

Specifically, if a habitual statistical analysis is performed on
any experimentally available set of frequency data, there is no
obvious method to extract a  reliable  information about
tendencies (local mean values) of the random dynamics.
Nonexistence of mean values and higher moments may also be
interpreted as the nonexistence of observable (e.g. mean, like
drifts or local currents) regularities of the dynamics. Moreover,
as we have learned before, the jump-type processes usually admit
arbitrarily small   jumps (with no lower bound) and finite, but
arbitrarily   large jump sizes (with no upper bound). Any
laboratory experiment or computer simulation would  involve both
the lower (coarse-graining) and upper  bound on the jump size.
Mathematically, that sets (as suitable) the framework of standard
jump processes for which the central limit theorem is known to
hold true in its Gaussian version (even if we account for a
possible abnormally slow convergence to a Gaussian). Therefore,
there is no clear-cut procedures allowing to attribute an
unambiguous statistical interpretation in terms of
 L\'{e}vy processes to  given  phenomenological data.
Moreover,  no realistic formulation of a
fluctuation-dissipation theorem is possible in that case (nonexistence
of variances) which pushes  us away from any conceivable thermal
equilibrium framework.

The starting point for Ornstein and Uhlenbeck (cf. the previous Section)
 was the dissipative Langevin equation
\be \frac{dV}{dt}\;=\;-\lambda V(t)\;+\;A(t)\ee where $V(t)$ is a
random variable describing the velocity of a particle,
$\lambda>0$ is a dissipation constant, and $A(t)$ is another
random variable whose probabilistic features  are determined by
the probability distribution of $V(t)$, which is assumed to
satisfy a concrete law when $t\to\infty$. Because $V(t)$ may have
no time derivative, the Langevin equation was soon replaced by
another one, the stochastic differential equation, namely \be
d{V}(t)\;=\;-\lambda V(t)dt\;+\; dB(t),\quad  V(0)\;=\;v_0\ee
which received a rigorous interpretation within the framework of
stochastic analysis. In the case when the probability
distribution of $V(t)$, $t\to\infty$, is the Maxwell one, $B(t)$
must be a Gaussian  (in fact Wiener, $B_t=W_t$) process, and then
we  end up with a classical Ornstein-Uhlenbeck process.

Now, we shall discuss properties of the process $V(t)$,
 in the case when $B\,=\,(B(t))_{t\geq 0}$ is the Cauchy process instead
 of the traditional Wiener one.
By straightforward integration  we obtain that for $t\geq s$ \be
V(t)\;=\;\els{V}(s)\;+\;\elt\cts e^{\lambda\tau} dB(\tau).\ee

There
are a number of (equivalent) procedures to deduce a probability
density of the process $ V(t)$ from the Cauchy increments
statistics. One may  follow a direct probabilistic route which,
upon assuming that a characteristic function of the Cauchy probability
measure reads
\be E[e^{ipB(t)}]\;=\;e^{-t\psi(p)}\ee
with the  choice of  $\psi(p)\,=\,\sigma^2|p|$ (before, we have
used  $\sigma ^2=1$),  leads to the transition density:
\be
p_{t-s}(u,v)=P[V(t)\,=\,u|V(s)\,=\,v]\;=\;\frac{1}{\pi}
\frac{\sigma^2(t\,-\,s)}{(u\,-\,v\els)^2\:+\:
\sigma^4(t\,-\,s)}\ee where $\sigma^2(t\,-\,s)\,=\,\sil
(1\:-\:\els)$.

Since $V(0)\,=\,v_0$,
the probability density of $V(t)$ is given by
$ P[V(t)\,=\,v]\;=\;\frac{1}{\pi}
\frac{\sit}{(v\:-\:v_0\elt)^2\:+\:\sik}$.

Now, we shall demonstrate  an important property (mentioned before
in connection with the Ornstein-Uhlenbeck process)
 of the so-called stochastic continuity which is a necessary
 condition
 to give a stochastic process an unambiguous status.
 Namely, we need to show that
 for any $\epsilon>0$ the following equation is satisfied
$ \lim\limits_{t\to s}P[|V(t)-V(s)| \geq\:\epsilon]=0$. This
equation is equivalent to $ \lim\limits_{t\to 0}\int\limits_{|u-v|
\geq\epsilon}p_t(u,v)du\;=\;0$.
Because of
$\int\limits_{|u-v|\geq\epsilon}p_t(u|v)du\:=\:1\:-\:\frac{1}{\pi}
[\arctan\frac{\epsilon\:+\:v(1\:-\:\elt)}{\sit}\:+
\:\arctan\frac{\epsilon\:-\:v(1\:-\:\elt)}{\sit}]$ and
remembering that $\sigma^2(t)\,=\,\sil (1\:-\:\elt)$, the
stochastic continuity property does follow.

The nonexistence   of moments of the probability measure in case
of the Cauchy process leads to straightforward  difficulties,
since the standard local characteristics of the diffusion-type
process like the
 drift and the diffusion function (or coefficient)   seem to be excluded
 in the present case.
However, for  the considered Ornstein-Uhlenbeck-Cauchy process,
the notion of the forward drift of the process proves to make
sense.

Since we know the Markov transition function
$p_{t-s}(u,v)$, $t\geq s$, for the process $V_t$, we can exploit
our experience with diffusion processes  and say that
the process $V_t$ has a drift (in fact, forward drift) if the
following limit
\be \lim\limits_{t\to s}\frac{1}{t-s}
\int\limits_{|u-v|\leq\delta }(u\:-\:v)p_{t-s}(u,v)du\ee
does not depend on the choice of $\delta >0$.
If so, then its value depending only on $(v,\,s)$ we
denote by $b(v,\,s)$ and call it the drift coefficient.
Clearly, if $p$ is homogeneous in time, then
the drift coefficient depends only on the  variable $v$.
Let us emphasize that in the above definition we do not
require the process $V_t$
 to have finite moments.\\
We claim that the jump-type Markov  process $V(t)$ has a (forward)
drift which reads  $b(v)\,=\,-\lambda v$. Indeed, by
 first evaluating  the indefinite integral
$$I\;=\;\frac{1}{\pi}\int (u\:-\:v) \frac{\sit
du}{(\yks)^2\:+\:\sik}$$ and  substituting $z\,=\,\yks$, we get
$$\frac{\sit}{\pi}\int\frac{zdz}{\zes}\;+\;
\frac{v}{\pi}(\elt\:-\:1)\int\frac{\sit dz}{\zes}$$
$$=\;\frac{\sit}{2\pi}\log (\zes)\;+\;
\frac{v}{\pi}(\elt\:-\:1)\arctan (\frac{z}{\sit}).$$ Hence
$$I\;=\;\frac{\sit}{2\pi}\log[(\yks)^2\:+\:\sik]\;+
\;\frac{v}{\pi}(\elt\:-\:1)\arctan[\frac{\yks}{\sit}]$$ and
consequently the limit
 $$\lit
I|^{u=v+\epsilon}_{u=v-\epsilon}\;=$$
$$\lit\frac{\sit}{2\pi}(\log[(v\:+
\:\epsilon\:-\:v\elt)^2\:+\:\sik]\;-\;
\log[(v\:-\:\epsilon\:-\:v\elt)^2\:+\:\sik])$$
$$+\;\lit\frac{v}{\pi}(\elt\:-\:1)(\arctan[\frac{v\:+
\:\epsilon\:-\:v\elt}{\sit}]\;-\;
\arctan[\frac{v\:-\:\epsilon\:-\:v\elt}{\sit}])$$
$$=\;0\;-\;\lambda\frac{v}{\pi}(\frac{\pi}{2}\:+
\:\frac{\pi}{2})\;=\;-\lambda v\quad $$ exists and is
$\epsilon $-independent. This is the forward drift of the process $V(t)$
which  proves a consistency of the derived transition probability
density with the stochastic differential equation for the process
$V_t=V(t)$.

It is well known that for Markovian diffusion processes all local
characteristics of motion (conditional expectation values that
yield drifts and variances)   are derivable from transition
probability densities, supplemented (if needed)
 by the density of the process.
We have demonstrated that, in the non-Gaussian  context,
 the nonexistence of moments
does not necessarily imply   the nonexistence of local
characteristics (drifts) of the process.

However, the situation becomes uncomfortable once we attempt to
evaluate another local moment.  Namely, in the present case there
holds: \be lim_{t\downarrow s} {1\over {t-s}} \int_{|v-u|\leq
\delta } (u-v)^2 p_t(u,v)du = {2\sigma ^2\over \pi } \delta \ee
i. e. an explicit cutoff $\delta $ (upper bound on the size of
jumps) persists in this formula and there is no way to remove
that jump size restriction from the formalism, unless we wish to
get the divergent  integral.

This property is  a clear indication that a convergence to a
Gaussian might always be expected if the Ornstein-Uhlenbeck-Cauchy
process is approximated (we disregard an issue of how good that
approximation is) by means of  jump processes with
 an upper and lower bound on the jump size. In that case both the
mean and variance would exist for the approximating process. In
particular, the central limit theorem would work as usual for the
$(\epsilon , \delta )$-jump process approximation  of
the Cauchy process $B_t$.\\

 {\bf
Acknowledgement}: The author receives financial support from the
KBN research grant No 2 P03B 086 16.

\end{document}